\documentclass[graybox]{svmult}
\usepackage{todonotes}
\usepackage{hyperref}
\usepackage{rotating}
\usepackage{comment}
\usepackage{adjustbox}
\usepackage{type1cm}        
\usepackage{makeidx}        
\usepackage{graphicx}   
\usepackage{multicol}        
\usepackage[bottom]{footmisc}
\usepackage{booktabs}
\usepackage{newtxtext}       
\usepackage{newtxmath}       

\makeindex             

\begin{document}

\title*{Graph Summarization}
\author{Angela Bonifati, Stefania Dumbrava and Haridimos Kondylakis}
 \authorrunning{A.Bonifati, S.Dumbrava and H. Kondylakis}
\institute{Angela Bonifati \at Lyon 1 University
\email{angela.bonifati @univ-lyon.fr}
\and Stefania Dumbrava \at ENSIIE \email{stefania.dumbrava@ensiie.fr}
\and Haridimos Kondylakis \at FORTH-ICS \email{kondylak@ics.forth.gr}
}

\maketitle

The continuous and rapid growth of highly interconnected datasets, which are both voluminous and complex, calls for the development of adequate processing and analytical techniques. One method for condensing and simplifying such datasets is \emph{graph summarization}. Its main goals are to \emph{reduce} graph data volumes, to \emph{accelerate} graph query evaluation, as well as to \emph{facilitate} graph visualization, analytics, and cleaning. \emph{Graph summarization} denotes a series of application-specific algorithms designed to transform graphs into more compact representations, while preserving structural patterns, query answers, or specific property distributions. As this problem is common to several areas studying graph topologies, different approaches, such as clustering, compression, sampling, or influence detection, have been proposed, primarily based on statistical and optimization methods. Despite these recent advances, the topic of graph summarization still presents open research challenges, in particular when dealing with richer graph models, such as property-based ones, when defining appropriate quality metrics, or when handling updates.

Up to date, there exist several surveys focusing on graph summarization. For example, a recent survey~\cite{DBLP:journals/csur/LiuSDK18} targets generic graph summarization methods, whereas in \cite{DBLP:journals/vldb/CebiricGKKMTZ19} and \cite{DBLP:conf/edbt/KondylakisKM19} approaches for summarizing semantic graphs are presented. Furthermore, in \cite{PAL19} graph-based methods for ontology summarization are discussed. Also, an outline of advances in graph partitioning/compression, as well as their applications to Big Data and the Semantic Web are given in \cite{S19} and \cite{BV19}, respectively.

The focus of our chapter is to pinpoint the main graph summarization methods, but especially to focus on the most recent approaches and novel research trends on this topic, not yet covered by previous surveys. In Section~\ref{sec:definition}, we introduce preliminary 
concepts.
In Section~\ref{sec:overview}, we discuss novel techniques for \emph{graph clustering}, as well as \emph{semantic}, \emph{dynamic}, and \emph{hybrid} graph summarization. In Section~\ref{sec:key_research_findings}, we structure these into a taxonomy and distil the key research findings 
targeting clustering, statistical, and goal-driven methods. In Section~\ref{sec:examples_of_application}, we highlight promising use-cases and applications and, in Section~\ref{sec:future_direction_for_research}, we discuss promising  future research directions.

\section{Preliminaries}
\label{sec:definition}

As the works we will discuss operate with different graph models, we start with a brief overview of useful notions. 

Let $\mathcal{O}$ be a set of \emph{objects} and $\mathcal{V} \subseteq \mathcal{O}$, a finite set of \emph{vertices}. An \emph{undirected graph} (UG) is a structure $(\mathcal{V}, \mathcal{E})$, s.t $\mathcal{E} \subseteq {\binom {\mathcal{V}} {2}}$ is a finite set of \emph{edges}. 
If $\mathcal{E} \subseteq {\mathcal{V} \times \mathcal{V}}$, $(\mathcal{V}, \mathcal{E})$ is a \emph{directed graph} (DG). We call both \emph{base graphs}. Base graphs s.t multiple edges connect the same two vertices are called \emph{multi-graphs}. Base graphs s.t 
an attribute list can be attached to each node/edge are called 
\emph{attributed graphs} (AG).
\begin{definition}[Knowledge Graphs] Given a set of RDF triples $\mathcal{T}$, a \emph{knowledge graph} (KG) is a multi-graph s.t 
$\mathcal{V} = \{s_i \mid (s_i,p_i,o_i) \in \mathcal{T} \} \cup \{o_i \mid (s_i,p_i,o_i) \in \mathcal{T} \}$ and $\mathcal{E} = \{p_i \mid(s_i,p_i,o_i) \in \mathcal{T} \}$. A special case is that of 
\emph{geographical knowledge graphs} (GG), where vertices can be mapped to meaningful geographic identifiers \cite{DBLP:phd/dnb/Yan19}.
\end{definition}
Let $\mathcal{L}$ be a finite set of \emph{labels} and $\lambda : \mathcal{V} \cup \mathcal{E} \rightarrow \mathcal{P}(\mathcal{L})$, a function assigning a \emph{finite set} of labels to each object. 
A \emph{labeled graph} (LG) is a structure $(\mathcal{V}, \mathcal{E}, \lambda)$. Further extending this class to \emph{directed, attributed multi-graphs}, we obtain the most expressive type of static graphs, i.e., \emph{property graphs}, defined below (see also \cite{BFVY18}).
\begin{definition}[Property Graphs] Let $\mathcal{K}$ be a set of \emph{property keys} and $\mathcal{N}$, a set of \emph{values}. A \emph{property graph} (PG) is a structure $(\mathcal{V}, \mathcal{E}, \eta, \lambda, \nu)$, where 
$\mathcal{V} \subseteq \mathcal{O}$ is a finite set of \emph{vertices},
$\mathcal{E} \subseteq \mathcal{O}$ is a finite set of \emph{edges}, 
$\eta : \mathcal{E} \rightarrow \mathcal{V} \times \mathcal{V}$ is a function 
assigning a pair of vertices to each edge, 
 and
$\lambda : (\mathcal{V} \cup \mathcal{E}) \times \mathcal{K} \rightarrow \mathcal{N}$
is a \emph{partial function} assigning property values to objects.
\end{definition}

In the \emph{dynamic} setting, the corresponding graph type for representing data streams \cite{HRR98} is given by \emph{streaming graphs} (SG) \cite{LVM18}.

\begin{definition}[Streaming Graphs] Let $\mathcal{T}$ be a set of timestamps. A \emph{streaming graph} is a structure 
$(\mathcal{T}, \mathcal{V}, \mathcal{W}, \mathcal{E})$, where
$\mathcal{W} \subseteq \mathcal{T} \times \mathcal{V}$ 
is a set of \emph{temporal nodes} and
$\mathcal{E} \subseteq \mathcal{T} \times {\binom {\mathcal{V}} {2}}$ 
is a set of \emph{links}, s.t $(t, \{u, v\}) \in \mathcal{E}$ implies
$(t, u) \in \mathcal{W}$ and $(t, v) \in \mathcal{W}$.
\end{definition}

\section{Recent Summarization Techniques}
\label{sec:overview}

We outline novel summarization techniques proposed in the literature. In Section \ref{subsec:topological}, we discuss \emph{graph clustering} methods, in Section \ref{subsec:statistical}, we present recent 
methods for \emph{statistical summarization}, whereas in Section \ref{subsec:goal_summaries}, we discuss \emph{goal-driven summarization} approaches for streaming and property graphs.

\subsection{Graph Clustering}
\label{subsec:topological}

Works in this category target \emph{graph clustering}-based approaches.
Graph clustering is one of the key techniques used in exploratory data analysis, as it allows to identify components that exhibit similar properties. In general, a graph cluster consists of nodes that are densely connected within a group and sparsely connected with outside ones. In order to understand the structure of large-scale graphs, it is important to not only compute such clusters, but to also identify the roles that the various nodes play within the graph. As such, nodes that bridge different clusters are distinguished as hubs and are considered to correspond to highly influential entities, while those that are neither clusters nor hubs are called outliers and are treated as noise. Such a differentiation is important when mining complex networks; for example, in web graphs, hubs link related pages, while outliers can correspond to spam. 

State of the art approaches to graph summarization through clustering roughly fall into two categories: \emph{structural} and \emph{attributed-based}. The following techniques operate of simple, undirected graphs, with the exception of the attribute-based ones, in which a list of feature attributes is also associated to each node.

\paragraph{\textbf{Structural Clustering.}}

\emph{Structural Clustering} takes into account the graph's connectivity and uses standard algorithms based on partitioning \cite{WXSW14} and on computing modularity, density, or custom measures, such as the reliable structural similarity introduced in \cite{QLLQWYM19} for clustering probabilistic graphs. Other approaches rely on identifying sets of k-median (respectively, k-center) nodes that maximize the average (respectively, minimum) connection probability between each node and its cluster's center \cite{WS11}. 

Structural clustering also often employs spectral methods, such as that of Laplacian eigenmaps to map nodes with higher similarity closer, based on a given symmetric, non-negative metric. However, a drawback of such techniques is that they are vulnerable to noise and outliers. To address this, recent techniques based on removing low-density nodes have been proposed. Recently, the work in \cite{KDK20} shows how to use a sparse regularization model, which reconstructs node density from a similarity matrix, to prune out the noise and detect clusters in the process.

Other structural methods factorize the node adjacency matrix to compute clusters (\cite{CLX15},\cite{NMV17}), low-dimensional node embeddings (\cite{CLX16}, \cite{WC016}, \cite{YCZ18}), or run random walks to learn such embeddings by maximizing neighbourhood probabilities (\cite{PAS14}, \cite{GL16}). In \cite{YBLLZ19}, a color-based random walk mechanism is presented, which allows identifying interactions between the seed nodes of local clusters.

The recent work in \cite{HGX19} extends k-median/k-center techniques to uncertain graphs and proposes several novel algorithms with provable performance bounds. In \cite{W0CL19}, an index-based algorithm is introduced for the structural clustering of undirected, unweighted graphs. The proposed methodology is based on the maintenance of structural similarity for each pair of adjacent vertices and is capable of handling updates. The work in \cite{LB20} addresses the problem of structural clustering, by using multi-scale community detection techniques based on continuous-nearest neighbours (CkNN) similarity graphs and Markov stability quality measures. Community detection techniques are used in \cite{KCC17} to compute graph clusters, while also taking into account node relevance with respect to given queries. To this end, the authors introduce the query-oriented normalized cut and cluster balance metrics and combine these to compute the output clustering.

The work of \cite{ZXZLZY19} frames graph clustering as an unconstrained convex optimization problem and proposes a technique to reorganize datasets into so-called triangle lassos, connecting similar nodes. A optimized, iterative version of the SCAN algorithm, anySCAN, is presented in \cite{MAABDJK19}, with the purpose of performing parallel clustering computations on large, static and dynamic graph datasets. In \cite{ZNCNZY19}, the machine learning technique of multi-view clustering is used to combine feature information from different graph views. These are then integrated into a global graph, whose structure is tuned through a specialized objective function ensuring that the number of components corresponds to that of clusters. In order to refine clustering results, game theoretical methods, based on consensus computation, have also been proposed. For example, in \cite{HAM19}, multiple graph clusters are integrated and outlier nodes are obtained through majority voting. 

Within the structural clustering category, one can distinguish between quotient and non-quotient approaches. On the one hand, \emph{quotient} methods are based on the notion of graph node "equivalence" and produce summaries by assigning a representative to each such equivalence class. A recent work in this area is given by \cite{DBLP:conf/edbt/GoasdoueGM19}, in which compact summaries of heterogeneous RDF graphs are built for visualization purposes. 
The approach ignores the schema triples, considering only type and data ones, and relies on the concept of property cliques, which encode transitive relations of edge co-occurrence on graph nodes.

The proposed algorithms are time linear in the size of the input graph and incremental. In \cite{DBLP:conf/www/ShinG0R19}, the authors present a fast summarization algorithm for graphs that are too large to fit in main memory, based on dividing these into smaller subgraphs, to be processed in parallel. Apart from the compact representation,
this summarization also produces edge corrections, allowing one to
restore the original graph, exactly or within given error bounds.

On the other hand, \emph{non-quotient} methods are usually based on centrality measures, selecting only specific graph subsets, as in \cite{DBLP:journals/ieicet/DingYZLG19}. This recent work proposes an algorithm enhancing topological summarization with semantic information. Thus, embeddings are generated to measure the extent to which concepts produce compact summaries, while similarity is captured by the distance between these embeddings. Next, k-means is used to select the important concepts and their similarity is further taken into account, in order
to avoid redundancy.

\paragraph{\textbf{Attributed Clustering.}}
\emph{Attributed Clustering} considers both the topology of the graph and a set of feature attributes that is attached to each node. To obtain consistent clusters, in this setting, nodes and features are either taken into account together, by matrix factorization and spectral clustering algorithms, or are integrated in graph convolutional networks (GCN) \cite{KW17}. In the latter case, a wide variety of graph auto-encoders (variational, marginal, adversarial, regularized) are then used to learn node representations and to reconstruct the adjacency matrix, as well as the different node features. Recently, the work of \cite{ZLLW19} has proposed to combine a high-order graph convolution method (for smooth feature representation) with spectral clustering on the learned features, to capture global structures and to adapt the convolution order to each dataset. Flow-based technique for local clustering are introduced in \cite{VKG19}, whereby semi-supervised information about target clusters is exploited to place constraints or penalties on excluding specific seed nodes from the output set. The underlying method in \cite{CGZJZ19} is based on a star-schema graph representation, in which attributes are modeled as different node types. DBSCAN clustering is then performed, using a personalized Pagerank as a unified distance measure for structural and attribute similarity.

\subsection{Statistical Summarization}
\label{subsec:statistical}
Statistical summarization mostly relies on occurrence counting and quantitative measures. Underlying approaches are based on either \textit{pattern-mining} or \textit{sampling}. Works in the former category aim to reveal patterns in the data and use these to summarize, while those in the later focus on selecting graph subsets.

The approach in \cite{DBLP:phd/dnb/Yan19} focuses on summarizing geographical knowledge graphs and introduces the concept of geo-spatial inductive bias
(knowledge patterns hidden within geographic components). It deals with
the summarization of both hierarchical and multimedia information related to the geographic nodes.

\subsection{Goal-Driven Summarization}
\label{subsec:goal_summaries}

In the above sections, we have discussed various methods for summarizing static graphs. However, many of the works focused on generating summaries are goal-driven
and set to optimize the memory footprint or some other utility type.

Following this direction, a key problem to tackle is that of summarizing dynamically changing graphs. These are graphs whose content (either edge labels, weights, or entire nodes and edges) is evolving over a sliding window of predefined size and are also known as \emph{graph streams under the window-based model} \cite{PBO20}. Such continuously changing graphs need to be summarized in a way that ensures the scalability and efficiency of the queries formulated on the obtained summary.   

\textit{Streaming graph summarization} approaches have recently appeared and leverage a common principle: the production of a concise representation that fits in memory.
Tsalouchidou et al. \cite{TsalouchidouBMB20} focus on the design of an online clustering algorithm that overcomes the basic stringent memory requirements of a baseline (based on $k$-means clustering \cite{RiondatoGB14}). They build on the micro-clusters concept from  \cite{AggarwalHWY03}, in order to provide a memory-efficient algorithm for continuously changing graphs. The idea is to leverage a time series of adjacency matrices, each of which represents a static graph. The latter can also be seen as an Order 3 tensor. The problem is then formulated in terms of tensor summarization, where a tensor summary is obtained for the last $w$ timestamps. Their distributed implementation allows dealing with large-scaled graphs on which temporal and probabilistic queries can be issued. 
The second approach \cite{Gou0Z019} also considers weighted graphs, where the weight is given by the timestamp, and strives to find an alternative data structure to the adjacency matrix, based on hash-based compression. In particular, a graph sketch, designed for sparse graphs, is created to store different source nodes/destination nodes in the same row/column and to distinguish them with fingerprints. The method outperforms the state of the art graph summarization algorithms, such as \cite{ZhaoAW11}, for most queries, including topological ones (such as reachability and successor queries). 

Other \textit{goal-driven summaries} have addressed the problem of creating query-aware, compact graph representations, starting from a weighted or a labeled graph instance. GRASP summaries \cite{DBDV19} have been defined for multi-labeled graphs that also possess node and edge properties. These knowledge-driven semantic graphs are also known as \emph{property graphs} (PGs). In GRASP, supernodes (superedges, resp.) are created to group together label-compatible graph nodes (edges, resp.), while also storing relevant statistical information. By incorporating this information, the obtained graph summaries are thus
tailored to highly accurate approximations of basic analytical queries. The target fragment is that of counting regular path queries, which allows one to estimate, for example, the number of connections established in a social network within a given period.
 
The second kind of summaries \cite{KumarE18} differ from the above in that they aim to maximize a utility function. While they also apply group-based iterative graph summarization as GRASP, their approach is not tailored to a specific query fragment. Contrary to GRASP, they also allow one to instantiate several utility functions, such as edge importance, edge submodularity, etc. In a sense, application-specific utility functions could
thus be encoded.

Furthermore, depending on the chosen utility function, their definition of error is different from GRASP and builds upon the reconstruction error, in case the graph summarization step is reverted. The high utility and scalability of their method is shown through a wide range of experiments. 
In addition, in \cite{DBLP:conf/icdm/SafaviBFMMK19}, the authors propose a personalized graph summarization method. The idea is to construct custom knowledge graph summaries, which only contain the most relevant information and which respect storage limitations. The problem is formalized as one of constructing a sparse graph that maximizes the inferred individual utility, subject to user- and device-specific constraints on the summary size.

\section{Key Research Findings}
\label{sec:key_research_findings}

\begin{figure}[hbt!]
 \centering
 \includegraphics[width=\linewidth]{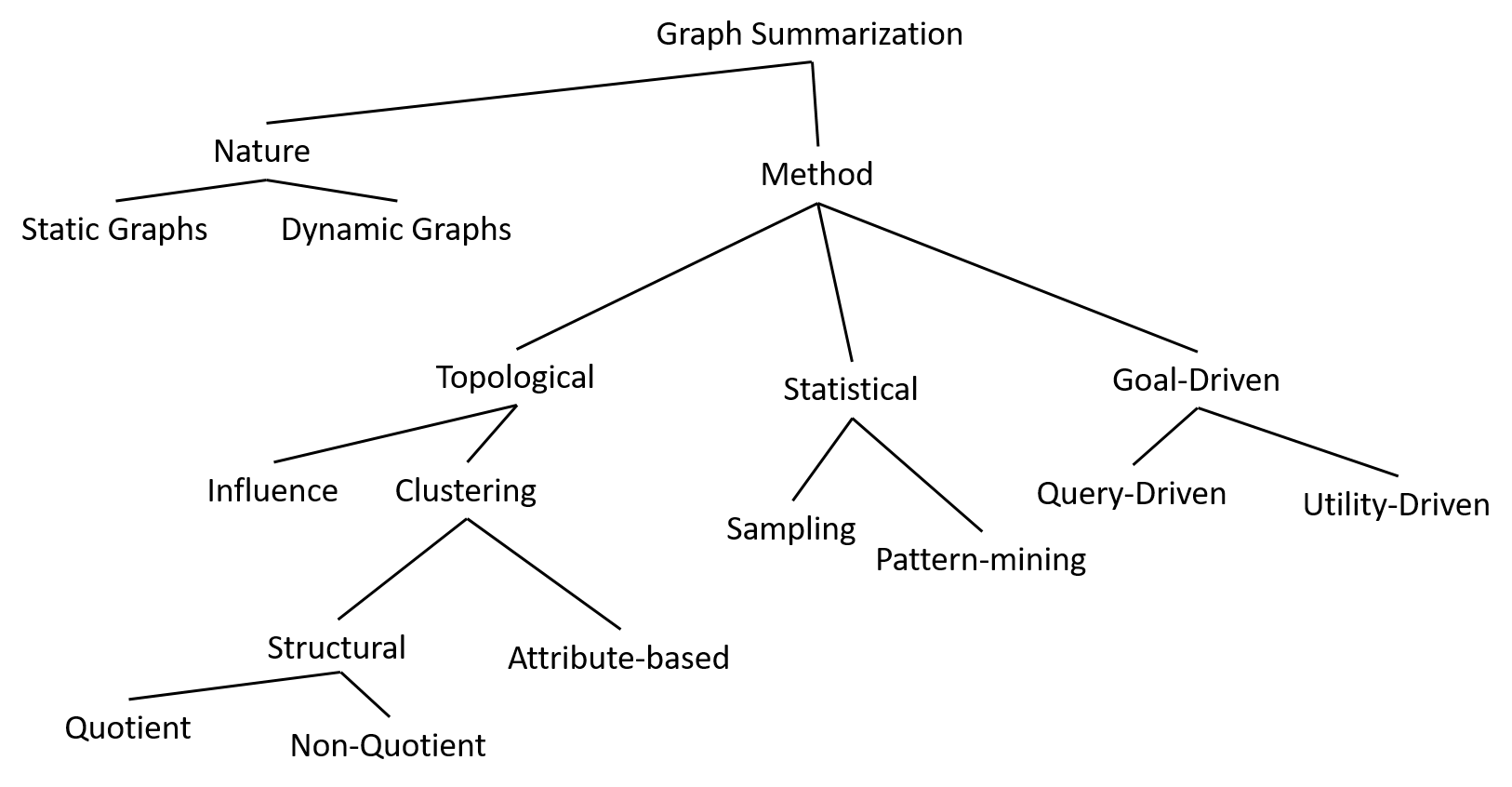}
  \caption{A taxonomy of the included works.}
  \label{fig:taxonomy}
\end{figure}

The summarization approaches discussed previously can be structured into the taxonomy depicted in Fig.\ref{fig:taxonomy}. We first notice that these methods apply to both static and dynamic graphs. Also, depending on their scope, the used techniques can be roughly classified as three-fold. First, those that rely on the underlying graph topology mainly perform clustering, by preserving structural or semantic (attribute-based) properties. Next, statistical means, ranging from sampling to complex pattern mining, are used to discover hidden information. Finally, goal driven approaches consider the relevance with respect to given queries or to pre-defined utility functions when summarizing.

We consider each of these directions and distil the topics currently in the limelight. 

Regarding \emph{graph clustering}, recent efforts focus on \emph{locality} and \emph{efficiency}. As such, \emph{flow-based algorithms} are adapted and improved, to render local clustering amenable to real-world, semi-supervised problems. Other methods target local clustering under constraints and employ \emph{colored random walks}, to account for prior knowledge. For efficiency purposes, \emph{index-based} approaches are used in structural clustering and tailored to efficient graph querying and index maintenance. Also, the challenging problem of \emph{uncertain graph} summarization has been recently tackled, by designing approximation algorithms with improved accuracy and performance. 

While most graph summaries are built through clustering techniques, we have seen that other approaches are also being successfully employed. For example, when considering quantitative criteria, statistical means can be used to extract relevant patterns. One recent application area is that of domain knowledge graphs, where geographic information can thus be compactly represented. Finally, utility functions, such as query relevance or memory footprints, can be taken into account when constructing summaries. This is especially relevant when dealing with expensive analytical queries, such as counting RPQs \cite{BD18}, or with large volumes of dynamic data, such as streaming graphs.

To better grasp the scope and purpose of the summarization approaches from Sec.~\ref{sec:overview}, we provide a 
classification in Fig.~\ref{fig:sum_table}. Note that the corresponding graph types are abbreviated, cf. Sec.~\ref{sec:definition}, as follows: \emph{undirected} (UG), \emph{labeled} (LG), \emph{attributed} (AG), as well as \emph{knowledge graphs} (KG), \emph{geographical graphs} (GG), \emph{property graphs} (PG), and \emph{stream graphs} (SG).

Inspecting the above table, we notice that most recent works have focused on structural clustering. While attributed approaches (\cite{ZLLW19}, \cite{CGZJZ19}) also take into account richer graph models, typically considering feature vectors associated to nodes, the full expressiveness of property graphs is only tackled in \cite{DBDV19}, for AQP summarization. 
\begin{center}
\begin{figure}[t!]
\resizebox{\textwidth}{!}{
\begin{tabular}{@{}l|l|l@{}}
\toprule
\textbf{Work/Method/Graph Type}                            & \textbf{Keywords}                 & \textbf{Purpose} \\ \midrule
\cite{QLLQWYM19} Similarity-based clustering (LG)
                 & Probabilistic graphs; Dynamic programming          
                 & Data mining \\
\cite{KDK20} Spectral clustering (UG)
             & Non-linear patterns; Density reconstruction; Node cutting  
             & Noise elimination  \\
\cite{YBLLZ19} Constrained local clustering (LG)                
           & Color-based random walk; Seed nodes  
           & Community detection             \\ 
\cite{W0CL19} Index-based clustering (UG) 
            & SCAN; Index maintenance; Core \& neighbour orders 
            & Querying            \\
\cite{LB20} Geometric-based clustering (LG)                    
           & Markov Stability; Similarity Graphs 
           & Community detection \\
\cite{KCC17} Query-oriented clustering (LG)                                     & Laplacian eigenmaps                               
           & Community detection  \\
\cite{ZXZLZY19} Convex clustering  (UG)                              
           & Triangle lasso; Unconstrained optimization; 
           Regularization                                           
           & Data analysis  \\
\cite{MAABDJK19} Anytime clustering (LG)                           
           & SCAN; Parallelization; Dynamic Graphs; Multicore CPU
           & Application-specific  \\
\cite{ZNCNZY19} Adaptive clustering (UG)
           & Multiview clustering and learning; Feature extraction
           & Unsupervised learning           \\
\cite{HAM19} Consensus clustering (UG)
           & Similarity graphs; Automatic partitioning               
           & Application-specific    \\
\cite{KW17}  Attributed clustering (LG)
           & Multi-layer graph convolutional network  
           & Semi-supervised learning        \\
\cite{ZLLW19} Attributed clustering (AG)    
            & Adaptive high-order convolution
            & Application-specific  \\
\cite{VKG19} Attributed clustering (UG)                                  & Flow-based local graph clustering
            & Community detection           \\
\cite{CGZJZ19} Attributed clustering (AG)                    
          & DBSCAN; Incrementality; Game theory          
          & Data Mining     \\
\cite{DBLP:conf/edbt/GoasdoueGM19} Structural quotient (KG)   &   Incremental; Property cliques                                                      & Visualization    \\
\cite{DBLP:conf/www/ShinG0R19} Structural quotient (UG)       &   Partitioning and Parallelization; Compression                                      & Querying                 \\
\cite{DBLP:journals/ieicet/DingYZLG19}  Structural non-quotient (KG)       &   Concept Vectors;  Structural and semantic embeddings   & Visualization                  \\
\cite{DBLP:conf/icdm/SafaviBFMMK19}  Structural non-quotient (KG)           & Personalization; Utility optimization  & Visualization    \\
\cite{DBLP:phd/dnb/Yan19}       Structural non-quotient  (GG)              &  Geospatial inductive bias; Hierarchical; Multimedia         & Visualization                 \\
                                       \cite{TsalouchidouBMB20}  Tensor summaries (SG) & Streaming graphs; Micro-clusters                                           & Querying\\                                              \cite{ZhaoAW11} Hash-based compression (LG) & Timestamped weighted graphs                                            & Querying
\\                                              \cite{DBDV19} Quotient summaries (PG) &   Property Graphs; Complex Path Queries                                         & Approx. Querying
\\                                              \cite{KumarE18} Utility-driven summaries (LG)  &                              Trade-off between error and utility              & Application-specific
\end{tabular}
}
\caption{Classifying Novel Summarization Approaches}
\label{fig:sum_table}
\end{figure}
\end{center}
\section{Applications}
\label{sec:examples_of_application}

In this section, we elaborate on potential use-cases for graph summaries.

\textbf{Query Efficiency. } As summaries are often compact representations of the original input graphs, they can be used as \emph{indexes} on the latter \cite{DBLP:journals/ws/KonrathGSS12}.
Consequently, for efficiency purposes, queries could first be formulated on the summaries. The obtained summary nodes could then further be matched with the nodes they represent.

\textbf{Query Size Estimation. } Summaries often include statistics about the original graph, which could be exploited to estimate the size of query results \cite{DBLP:journals/tkde/LeLKD14}.

\textbf{Query Disambiguation. } Queries that contain path expressions with wildcards  are difficult to evaluate, despite being common in practice. A summary can easily provide information on the connectivity of the initial nodes and, as such, enable
queries to be more efficiently evaluated via rewriting \cite{DBLP:conf/vldb/GoldmanW97}.

\textbf{Source Selection. } Another interesting application is the use of summaries to detect whether a graph is likely to have specific information of potential interest for the user, without actually having to inspect the real data source \cite{LW17}. 

\textbf{Graph Visualization. } An obvious application for summaries is to enable the exploration of the original data source, effectively reducing the number of nodes/edges to be perceived by the user (\cite{DS13}, \cite{KKVF14}, \cite{DBLP:conf/semweb/TroullinouKSP18}, \cite{DBLP:conf/semweb/TroullinouKSP18a}, \cite{DBLP:conf/esws/PappasTRKP17}).

\textbf{Schema Discovery. } When no schema is present in the initial graph, a summary can be used instead to help users understand the original content, as shown in \cite{BKLK18}.

\textbf{Pattern Extraction. } Summarization also enables pattern identification and extraction \cite{KKVF14}, by abstracting away irrelevant graph portions. An interesting such use-case is given by blockchain-based crypto-currencies. In this setting, transactions correspond to openly-accessible graphs, whose topological features can shed light on the role and interactions of the participants. Graph analysis techniques can be thus applied to identify salient structural patterns.

\textbf{Knowledge Graph Search.} Specialized summaries \cite{SongWLDS18} can drive the search strategy in knowledge graphs. These represent lossy replacements of complex graph pattern and can be directly queried as approximate graph materialized views.  

\section{Future Research Directions}
\label{sec:future_direction_for_research}

In this section, we discuss future directions for graph summarization, as inspired by the existing literature. 

In the area of graph clustering, further improvements are needed to cope with \emph{mixed datasets}, in which data points are comprised of both numerical and categorical attributes. For such datasets, one has to design custom models, capable of handling missing or uncertain feature values, as well as explainable and interpretable clustering algorithms. Explainability of clustering results would also be beneficial for graph summarization, in order to tune results to particular use cases. 

The problem of building overlapping graph clusters, as addressed by \emph{fuzzy clustering} algorithms, is also interesting to consider and its implications for graph summarization are tangible. 

Moreover, we note that most existing approaches build static summaries. However, the used input graphs are constantly evolving and being updated. To address this, new research is tackling the problem of \textit{dynamicity}. As summary recomputation is often costly, novel insights are needed on how to efficiently achieve \textit{incrementality}. On a related note, recent works also focus on \textit{streaming graphs}, as summarization techniques are required to handle the constantly arriving flow of data that cannot actually be stored. In this setting, ensuring that streaming summaries are updatable, for example using a sliding windows approach, is essential for efficient processing. 

Furthermore, another interesting future direction would be to investigate \textit{quality metrics for summaries and evaluation benchmarks}. However, as graph summarization employs numerous techniques, different outputs might be produced, depending on the purpose, rendering the task difficult. 

Finally, the problem of graph summarization has been extensively addressed for existing graph data models, such as RDF, labeled, and weighted graphs. However, principled approaches would be desirable for more expressive graph data models, such as property graphs. On these graphs, clustering, in particular attributed methods also using edge features, dynamicity and benchmarking are all viable future research directions to be pursued.  

\bibliography{bibliography}
\bibliographystyle{spmpsci} 

\end{document}